\documentclass[prl,aps,graphicx,twocolumn,floats,bm]{revtex4}
\usepackage{graphicx}
\begin{document}


\title{Big, Fast Vortices in the d-RVB theory of high temperature superconductivity}
\author{L. B. Ioffe$^{1}$ and A. J. Millis$^{2}$}

\address{$^{1}$Center for Materials Theory \\
Department of Physics and Astronomy, Rutgers University
\\136 Frelinghuysen Rd, Piscataway NJ 08854\\
${^2}$Department of Physics and Astronomy\\
Columbia University, 538 W 120th St, NY Ny 10027}

\date{Feb 9, 2002}

\begin{abstract}
The effect of proximity to a Mott insulating phase on the superflow
properties of a d-wave superconductor is studied using the slave boson-U(1)
gauge theory model. The model has two limits corresponding to
superconductivity emerging either out of a 'renormalized fermi liquid' or
out of a non-fermi-liquid regime. Three crucial physical parameters are
identified: the size of the vortex \textit{as determined from the
supercurrent it induces;} the coupling of the superflow to the
quasiparticles and the 'nondissipative time derivative' term. As the Mott
phase is approached, the core size as defined from the supercurrent
diverges, the coupling between superflow and quasiparticles vanishes, and
the magnitude of the nondissipative time derivative dramatically increases. The
dissipation due to a moving vortex is found to vary as the third power of
the doping. The upper critical field and the size of the critical regime in
which paraconductivity may be observed are estimated, and found to be controlled
by the supercurrent length scale.
\end{abstract}
\vskip1pc]
\maketitle
\bigskip 


\section{\protect\bigskip Introduction}

High-$T_{c}$ superconductors are created by doping an antiferromagnetic
'Mott insulating' parent material, and the effect of proximity to the Mott
phase on their superconducting properties remains a crucial and still
incompletely understood issue \cite{Orenstein00}.\ One expects on general
grounds that the suppression of current response near a Mott insulator leads
to 'type II' behavior, so a fundamental issue is the physics associated with
vortices in the superconducting order parameter. An isolated vortex involves
a quantized flux ($hc/2e$ in conventional superconductors), a circulating
supercurrent pattern and a 'core region' in which the quasiparticle
excitation spectrum differs from that observed far from the vortex. The
possibility (apparently not realized in known superconductors) that
proximity to the Mott phase could induce an unconventional value of the flux
quantum has been discussed \cite{Sachdev92,Moler01}. An extensive literature
exists on quasiparticle properties (including the possibility of interesting
discrete core states \cite{quasiparticlerefs}, and whether an
antiferromagnetic \cite{Arovas97,Han00} or other \cite{Kishine01} state is
induced in or near the vortex core). However, apart from the pioneering
phenomenological work of Lee and Wen \cite{Lee97} and an analysis of the
resistive transition in overdoped $Tl$-based materials \cite{Geshkenbein98},
little theoeretical attention has been paid to the superflow properties even
though these in fact control many physically important quantities including $%
H_{c2}$ and the size of the 'critical regime' in which superconducting
fluctuation properties may be observed in the conductivity.

Theoretical analysis of vortex properties requires a model. Conventional
models of superconductivity in interacting electron systems are based on
Landau's fermi liquid theory, but as we show in the Appendix, analysis of
the changes occurring as the Mott phase is approached requires a model which
goes beyond Fermi liquid theory, at minimum by including effects
corresponding to a scale dependence of a Landau parameter and perhaps more
fundamentally by allowing for superconductivity to emerge from a
fundamentally non-fermi-liquid state or regime. One widely studied
theoretical model of a doped Mott insulator is the $U(1)$ gauge theory
implementation \cite{Ioffe89} of the RVB ideas of P. W. Anderson\ \cite
{Anderson87}. \ This theory and its variants have been extensively studied
as an approximation to the low energy physics of the $t-J$ model believed 
\cite{Orenstein00} to capture the essential aspects of the low energy
physics of high-T$_{c}$ materials. It exhibits (at least in a large-N limit)
a non-fermi-liquid regime \cite{Ioffe89,Nagaosa90} involving exotic
excitations ('spinons' and 'holons' coupled by a gauge field) and a fermi
liquid regime in which the spinon and holon are bound together into a
conventional electron and the gauge field a effects produce a nontrivial
doping dependence of the Landau parameter $F_{1S}$ \cite{Grilli90,Millis96}.
The model also possesses a d-wave superconducting state \cite{Kotliar88}
which may emerge either from the fermi-liquid or non-fermi-liquid regimes.
Quasiparticle properties (including a possible antiferromagnetic \cite{Han00}
or staggered flux \cite{Kishine01} state in the core of the vortex) have
been studied and the model has been shown to admit $h/2e$ vortices \cite
{Lee00}, but superfluid properties such as the supercurrent distribution in
the vortex state and the dissipation occurring when a vortex moves have been
less well studied.

As discussed at length elsewhere \cite{Lee97,Millis96,Ioffe01a} this theory
disagrees with experiment in a number of ways. Most problematically, the
model predicts a strong doping dependence to the leading low-T correction to
the London penetration depth ($d\lambda ^{-2}/dT\sim (doping)^{2}$) which is
not observed. We therefore do not believe the theory is a realistic
representation of high temperature superconductors; however it is a very
useful model system. We stress that although as usually formulated the model
involves exotic excitations such as holons and spinons, for the properties
we discuss these can be completely eliminated: as shown in the text and
Appendix, the model can be viewed simply as a method of calculating the
behavior of a fermi-liquid based system at length scales short enough that
scale dependence of the Landau parameters becomes important.

The importance of the model is that it provides an explicit realization of a
situation (which, we believe, is generically realized in lightly doped Mott
insulators) where the supercurrent-defined and quasiparticle-defined length
scales are parametrically different. We show in this paper that the longer
length scale is in fact the one relevant to the conventional superfluid
properties such as $T_{c}(H)$ and the size of the 'fluctuation regime' in
which the model exhibits (for example) a nontrivial paraconductivity.

The rest of this paper is organized as follows. Section II reviews necessary
aspects of the $U(1)$ gauge theory formalism. Section III discusses in
detail the current distribution around a vortex. Section IV calculates the
dissipation induced when a vortex moves and uses this information to
estimate the size of the 'critical regime' in which superconducting
fluctuation effects are visible in the conductivity. Section V considers the
low temperature limit of the upper critical field. Section VI is a
conclusion, summarizing the results and their implications. An Appendix
explicates the relation between the results presented here and the
conventional fermi liquid analysis.

\section{Formalism}

This subsection reviews results obtained in the early days of the gauge
theory \cite{Ioffe89,Kotliar88,Nagaosa90,Lee00}, in order to establish
notation and introduce important concepts.

In the gauge theory one writes the electron $c_{i\alpha }$ in terms of a
charge-e boson $b$ (representing a hole) and a fermionic 'spinon'
representing a spin degree of freedom, thus $c_{i\sigma
}^{+}=b_{i}f_{i\sigma }^{+}$. The superconducting state is described by a
d-wave BCS pairing of spinons \cite{Kotliar88} (involving a d-symmetry
pairing gap \ with maximum value $\Delta $) and a condensation of the
bosons. The low energy, long-wavelength physics is controlled by the
Hamiltonian 
\begin{eqnarray}
H_{gauge} &=&\frac{1}{2}\rho _{B}(\nabla \phi _{B}-a-A)^{2}+\frac{1}{2}\rho
_{F}(\frac{1}{2}\nabla \phi _{F}-a)^{2}+  \label{Lioffe} \\
&&+H_{D}+H_{mix} +...  \nonumber
\end{eqnarray}
Here $\ a$ is an internal gauge field which enforces the constraint, arising
because the physical fermion $c_{i}=f_{i}b_{i}^{+},$ that a \ longitudinal
spinon current must cause an equal and opposite boson current. (It is
possible to have transverse currents of spinons with no holon motion but
these are not relevant here). \ $\phi _{B}$ is the phase of the boson field
and $\rho _{B}$ is the $T=0$ boson superfluid stiffness, $\phi _{F}$ is the
phase of the spinon (fermion) pairing amplitude and $\rho _{F}$ is the
corresponding $T=0$ spinon 'superfluid' stiffness. $H_{D}$ is the usual
(normal-ordered) dirac Hamiltonian describing the quasiparticle part of the
spinon degrees of freedom and the ellipsis expresses terms irrelevant to the
present discussion.

$H_{D}$ has eigenvalues $E_{p}=\sqrt{v^{2}p_{1}^{2}+v_{2}p_{2}^{2}}$ with $%
v_{1}$ the spinon fermi velocity and $v_{2}$ related to the d-wave gap in
the usual way. The spinons are coupled to the gauge field and thus to the
'holons' via the term 
\begin{equation}
H_{mix}=\sum_{\alpha ,\sigma }\left( i\frac{1}{2}\partial _{\mu }\phi
_{F}-ea_{\mu }\right) \cdot \overrightarrow{v}_{1}c_{\alpha \sigma
}^{+}c_{\alpha \sigma }  \label{Hmixspinon}
\end{equation}

We begin our analysis of $H_{gauge}$ by considering length and energy
scales. The fermionic part of the Hamiltonian involves the length scale 
\begin{equation}
\xi _{F}=\frac{v_{1}}{\Delta }  \label{xiF}
\end{equation}
and two energy scales: $\Delta $ and 
\begin{equation}
\rho _{F}\sim v_{1}p_{F}  \label{rhospinon}
\end{equation}
$\xi _{F}$ is relatively short and does not diverge as the Mott phase is
approached, and $\rho _{F}$ is relatively large \ (of order $J$ in the $t-J$
model) and does not vanish as the Mott phase is approached.

The boson stiffness $\rho _{B}$ has dimension of energy (in two spatial
dimensions) and is proportional to the doping $x$ and to the basic
electronic hopping parameter $t$. $\rho _{B}$ is expected to vary \cite
{Popov} on the length scale $x^{-1/2}$ which is the distance between charge
carriers. \ We shall be interested primarily in the limit $x^{-1/2}>\xi _{F}$%
. \ We note that as the doping is increased, $\Delta $ decreases \cite
{Kotliar88} and eventually becomes smaller than an energy of order $x^{1/2}J$
so the inequality is reversed. For larger dopings the theory becomes
essentially the familiar $BCS$ one, with only one important length scale, $%
\xi _{F}$.

The currents carried by boson and fermion degrees of freedom are,
respectively 
\begin{eqnarray}
j_{B} &=&\rho _{B}\ast (\nabla \phi _{B}-a-A)  \label{jb} \\
j_{F} &=&\rho _{F}(\frac{1}{2}\nabla \phi _{F}-a)  \label{jf}
\end{eqnarray}
Here the $\ast $ denotes convolution and is to remind the reader that $\rho
_{B}$ is scale dependent on scales relevant to the subsequent discussion.

The physical current $j_{phys}=j_{B}$ and the constraint enforced by the
gauge field $a$ is $j_{B}+j_{F}=0$, i.e. 
\begin{equation}
\rho _{B}\ast (\nabla \phi _{B}-a-A)+\rho _{F}(\frac{1}{2}\nabla \phi
_{F}-a)=0  \label{constraint}
\end{equation}
This implies 
\begin{equation}
a=\left( \rho _{B}+\rho _{F}\right) ^{-1}\ast \left( \rho _{B}\ast (\nabla
\phi _{B}-A)+\rho _{F}\nabla \phi _{F}\right)  \label{a}
\end{equation}
In the long wavelength limit the non-locality of $\rho _{B}$ may be
neglected. As $T\rightarrow 0$ and assuming no fermions are excited,
elimination of $a$ leads to \cite{Ioffe89,Lee00} 
\begin{equation}
H_{phase}=\frac{\rho _{B}\rho _{F}}{\rho _{B}+\rho _{F}}\left( \nabla \phi
_{B}-\frac{1}{2}\nabla \phi _{F}-A\right) ^{2}  \label{Hphase}
\end{equation}
The meaning of this equation is that in a state with paired spinons and
condensed bosons at long wavelengths only the combination $\nabla \phi _{B}-%
\frac{1}{2}\nabla \phi _{F}$ couples to an external vector potential or is
relevant to the energy, and the physical superfluid stiffness $\rho _{S}=%
\frac{\rho _{B}\rho _{F}}{\rho _{B}+\rho _{F}}$. Similarly one finds 
\begin{equation}
H_{mix}=\sum_{\alpha ,\sigma }\frac{-\rho _{B}}{\rho _{B}+\rho _{F}}\left(
\nabla \phi _{B}-\frac{1}{2}\nabla \phi _{F}-A\right) \cdot \overrightarrow{v%
}_{1}c_{\alpha \sigma }^{+}c_{\alpha \sigma }  \label{Hmixspinon2}
\end{equation}
Because the gauge field has been eliminated, the fermionic degrees of
freedom should be regarded not as spinons but as Bogoliubov quasiparticles
of the superfluid (or near superfluid) state. They couple only to the
combination $\nabla \phi _{B}-\frac{1}{2}\nabla \phi _{F}$ and the coupling
is via an effective charge $Z^{e}=\frac{-\rho _{B}}{\rho _{B}+\rho _{F}}$
which is negative (hole-like) and vanishes as the Mott insulator is
approached.

Eqs \ref{Hphase},\ref{Hmixspinon2} constitute a derivation, from the $U(1)$
gauge theory, of phenomenological equations discussed in \cite{Ioffe01a}.
The derivation makes it clear that deviations from phenomenological theory
occur at length scales shorter than that specified by the scale dependence
of the physical superfluid stiffness, i.e. than the shorter of $\xi _{F}$
and $x^{-1/2}$. The derivation also makes it manifest that the
phenomenological action, discussed in \cite{Millis96} on the basis of fermi
liquid theory, is more general, and may apply also to situations in which
the normal state is not described by fermi liquid theory.

\section{Vortex solution--static case}

Consider a vortex. \ Far from the vortex core the fields are found by
minimizing $H_{phase}$ (Eq \ref{Hphase}) which implies that $\nabla
^{2}\left( \phi _{B}-\frac{1}{2}\phi _{F}-A\right) =0$. Single-valuedness of
the wave function implies that both $\phi _{B}$ and $\phi _{F}$ must have
circulation which is an integer multiple of $2\pi $ so that in a mean field
approximation one would write 
\begin{eqnarray}
\nabla \phi _{B} &=&\frac{m\widehat{\theta }}{r}  \label{phibvortex} \\
\nabla \phi _{F} &=&\frac{n\widehat{\theta }}{r}  \label{phisvortex}
\end{eqnarray}
The energy associated with a vortex is thus, approximately, 
\begin{equation}
E_{V}=E_{core}(n,m)+\frac{1}{2}\rho _{S}\left( m-\frac{1}{2}n-A\right)
^{2}\ln \left( \frac{R}{\xi }\right)  \label{Ev}
\end{equation}
where $\rho _{S}$ is the physical superfluid stiffness defined below Eq \ref
{Hphase}, $R$ is of the order of the inter-vortex separation and $\xi $ is
the length scale below which the supercurrent magnitude deviates from $1/r$
and will be discussed more fully below. $E_{core}$ is the core energy of the
vortex, i.e. the contribution to the energy arising from scales less than $%
\xi $.

The superflow contribution is clearly minimized by the choice $m=0$, $n=1$,
corresponding to a conventional $h/2e$ vortex. The core energy term requires
more discussion. If $n=1$ then a singularity in the fermion pairing
amplitude is required. In a clean conventional superconductor one would
estimate the energy cost of this singularity as the product of the
condensation energy per unit area ($N_{0}\Delta ^{2}$ with $N_{0}$ the
density of states) and the area of the core ($\xi _{F}^{2}=v_{F}^{2}/\Delta
^{2}$) leading to $E_{core}\sim v_{F}^{2}N_{0}$. In the present problem this
implies an $E_{core}$ of the order of the effective fermi energy $J$. On the
other hand, if $n=0$ and $m=1$ then no singularity is required in the
fermion field and a calculation very similar to that given in Eqs \ref
{boseschr} then shows that the core energy is of the order of the boson or
superflow energy $xt$, and can be absorbed into the definition of $\xi $.
These considerations suggest that in the gauge theory the vortex energy may
be estimated by 
\begin{equation}
E_{V}\approx C_{core}J\left( 1-\delta _{n,0}\right) +C_{sf}xt\ln \left( 
\frac{R}{\xi }\right) \left( m-\frac{1}{2}n-A\right) ^{2}  \label{Ev2}
\end{equation}
with $C_{core,sf}$ constants. This estimate (proposed and presented in more
sophisticated form by Sachdev \cite{Sachdev92}) suggests that when the
superflow energy is dominant (low vortex density or high doping) one has
conventional $h/2e$ vortices but that as $x$ is reduced or $R$ is decreased
a transition to doubly quantized vortices may occur.

This argument, however, is vitiated by recent work on the structure of the
vortex core. From different points of view the authors of \cite
{Arovas97,Han00,Kishine01} show that (within certain reasonable assumptions)
some other ordered state, also characterized by an electronic gap of the
order of $\Delta $, is very nearby in energy and indeed becomes favored as $%
x\rightarrow 0$. The consequence is that $C_{core}$ in Eq \ref{Ev2}
decreases rapidly as $x\rightarrow 0$ and indeed may even become negative,
implying that conventional ($h/2e$) vortices are always favored. In more
physical terms, within the classes of models (including the gauge model)
considered by \cite{Arovas97,Han00,Kishine01} the reason that the ground
state has superconducting rather than some other sort of order is the gain
in energy associated with establishing superfluid phase coherence, so it is
natural that even the vortex core energy is set by the phase stiffness. The
conventional nature of vortices in this theory was stressed in \cite{Lee00}
which however did not consider the core energy explicitly. We note however
that even if not favored in the superfluid state, doubly quantized $(h/e)$
vortices may be easily excited thermally once the long-ranged superfluid
order is disrupted.

We now study the structure of the vortex at shorter length scales. We
consider the case of very weak applied field, so we may choose a gauge in
which $A=0$. We assume $\rho _{B}<<\rho _{F}$ and expand in powers of $\rho
_{B}$. We find from Eqs \ref{a}, \ref{phibvortex} 
\begin{equation}
a=\frac{1}{2}\nabla \phi _{F}-\frac{1}{4\rho _{F}}\rho _{B}\ast \nabla \phi
_{F}+...  \label{avortex}
\end{equation}
where the ellipsis denotes both terms higher order in $\rho _{B}/\rho _{F}$
and fluctuations about the mean field solution for $\phi _{F,B}$.

We must now determine the behavior of the boson field and the fermion
pairing amplitude. The fermion pairing amplitude varies on the scale $\xi
_{F}$ which by assumption remains finite as the Mott phase is approached,
whereas one expects the bose amplitude to vary on the scale set by the
spacing between carriers, which diverges as the Mott phase is approached. \
We therefore focus on the bose field. \ In a lightly doped Mott insulator
the density of bosons is low. In the limit of dilute bosons one expects \cite
{Popov} that the bose amplitude is described by the two-dimensional
Hamiltonian density 
\begin{equation}
H_{bose}=\frac{1}{2m_{B}}\left( \left( \nabla -a-A\right) \psi \right) ^{2}+%
\frac{1}{2}\mu \psi ^{2}+\frac{1}{4}U\psi ^{4}  \label{Lbose}
\end{equation}
In the dilute limit, the parameters $\mu $ and $U$ are universal, given in
terms of the boson density $x$ and mass $m_{B}$ by 
\begin{eqnarray}
\mu &=&Ux  \label{muuniv} \\
U &=&\frac{4\pi }{m_{B}\ln (1/x)}  \label{Uuniv}
\end{eqnarray}
We expect that $m_{B}\sim 1/tb^{2}$ with $b$ the underlying lattice
constant. Also, the dilute limit means that a mean field approximation for
the boson field is reliable.

We must however consider the mean field approximation for the gauge field in
more detail. In the d-wave $RVB$ state, fluctuations in $a$ are controlled
by the stiffness corresponding to fermion pairing. This stiffness is large
at length scales longer than $\xi _{F}$ or energy scales less than the
fermion pairing amplitude $\Delta $ so a mean field approximation is
expected to be reliable at long length and low energy scales. However, if $%
\xi _{F},\Delta ^{-1}$ are shorter than the relevant bosonic length scale $%
x^{-1/2}$ and time scale $\left( xt\right) ^{-1}$ then fluctuations in $a$
may appreciably renormalize the parameters in $L_{bose}$. \ Because the
decay with $q$ of the fermionic stiffness is slow ($\sim \left( \xi
_{F}q\right) ^{-1}$) we focus here on the energy scale. \ As \ carriers are
added to a lightly doped Mott insulator, $\Delta $ decreases and $xt$
increases. When $\Delta /xt$ becomes less than unity we expect that
fluctuation corrections to the various parameters become large. We thus
distinguish two regimes: a ''super-dilute regime'' in which $\Delta >xt$ and 
$\gamma \sim 1$ and a ''dilute boson regime'' in which $x<<1$ but 
$\Delta<xt $ and $\gamma <<1$.

In the limit of interest $\xi _{F}<x^{-1/2}$ the boson ground state in the
presence of a vector potential is given by the solution of 
\begin{equation}
\frac{-1}{2m}\left( \nabla +a+A\right) ^{2}\psi +U\psi ^{3}=\mu \psi
\label{boseschr}
\end{equation}
Eqs \ref{boseschr},\ref{phisvortex},\ref{avortex} imply that in radial
coordinates we have (up to terms of relative order $x$) 
\begin{equation}
\frac{-1}{2m}\left( \partial _{r}^{2}+\frac{1}{r}\partial _{r}+\frac{1}{%
4r^{2}}\right) \psi +U\psi ^{3}=\mu \psi
\end{equation}
Defining 
\begin{equation}
\psi =x^{1/2}f(r/\xi )
\end{equation}
with 
\begin{equation}
\xi ^{-2}=2m\mu
\end{equation}
leads to the solution shown in Fig 1. \ In particular, at 
large distance $f\rightarrow 1$ while at small distance 
\begin{equation}
f=\sqrt{r}f_{0}
\end{equation}
with $f_{0}=0.886$. In other words, as the core of the vortex is
approached, the bose amplitude decreases as the square root of the distance
from the vortex core. 

\begin{figure}[ht]
\includegraphics[width=3.0in]{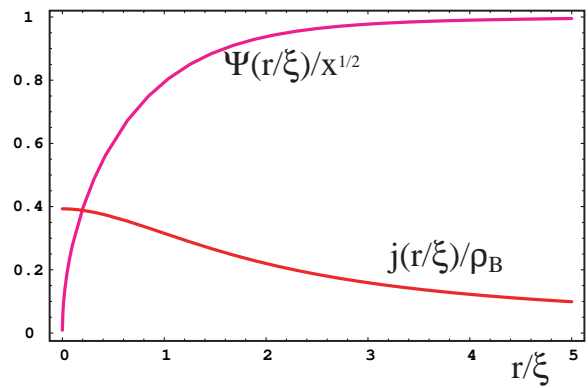}
\caption{Variation with distance from vortex core of boson amplitude
$\psi $ and supercurrent $j$}
\end{figure}

The supercurrent is given from Eqs \ref{jb},\ref{Lbose}  as 
\begin{equation}
j(r)=\frac{\left| \psi (r)\right| ^{2}}{m_{B}}a
\end{equation}
so in particular at small distances, in physical units 
\begin{equation}
j=\frac{f_{0}}{2\xi }\rho _{B}(\infty )
\end{equation}
The resulting current profile is also shown in Fig 1; we see that the
supercurrent varies as $1/r$ for $r>x^{-1/2}$ and is constant for smaller $r$%
, justifiying the qualitative statements made in \cite{Lee00}.

The fermion spectrum retains its long-distance value down to a length $\xi
_{F}=v_{1}/\Delta $ which is parametrically less than $\xi $ as $%
x\rightarrow 0$; below this length a variety of interesting physical effects 
\cite{Han00,Kishine01} may occur. The physical electron spectrum, observable
(in principle) via tunnelling, is calculated in the $U(1)$ gauge theory as a
convolution of a holon and a spinon \cite{Anderson87}, and so is relatively
broad in the non-superconducting phase of the model. In the superconducting
state the $q=0$ boson amplitude develops an expectation value and so the
electron spectral function acquires a sharp 'quasiparticle pole' feature. In
the limit $x\rightarrow 0$ the short length scales of the spinon spectrum
control the convolution so the coherent part of the spectrum at distance $r$
from the vortex core is proportional to the boson amplitude at distance $r$.
In other words, in this theory the strength of the quasiparticle peak
measured at a distance $r$ from the vortex core should begin to decrease for
\ as $r$ is reduced below $\xi $. This effect is not visible in published
tunnelling data \cite{Pan00}.

To summarize, in the $U(1)$ gauge theory of a lightly doped Mott insulator,
a vortex is characterized by two length scales: $\xi \sim x^{-1/2}$, below
which the supercurrent ceases to vary as $1/r$ (and in fact becomes
essentially $r$-independent) and the scale $\xi _{F}$ which does not diverge
as the Mott phase is approached and which controls the quasiparticle
properties. The core energy is small (of order the superfluid stiffness);
however the state in the core possesses a gap very similar to the
superconducting gap. This behavior should be contrasted with that of a
conventional (BCS) superconductor, in which the length defined by the
supercurrent is essentially the same as the length defined by the
quasiparticle properties and the core energy is large (of the order of the
fermi energy) and the core is gapless apart from the 'finite size effects'
which lead to the Caroli-Matricon states.

\section{Moving vortex}

We consider a slowly moving vortex with center position $\overrightarrow{X}%
_{v}(t)$, so that time derivatives of fields may be replaced by the dot
product of a field gradient and the vortex velocity; for example $\partial
_{t}\psi (r,t)=\partial _{t}\overrightarrow{X}_{v}(t)\cdot \overrightarrow{%
\nabla }\psi $. The contribution to the action from vortex motion has two
terms: a non-dissipative term corresponding to motion in an effective
magnetic field $B_{eff}$ and a dissipative term arising because vortex
motion excites fermionic excitations. These terms imply a classical equation
of motion 
\begin{equation}
B_{eff}\widehat{z}\times \partial _{t}X_{V}+\eta \partial _{t}X_{V}=F_{V}
\label{xv}
\end{equation}
where $F_{V}$ represents the forces acting on the vortex (arising for
example from an imposed current and from vortex-vortex interactions). Eq \ref
{xv} applies only for frequencies less than a cutoff frequency which is the
minimum of $\Delta $ and the boson frequency scale $xt$.

$B_{eff}$ may be obtained by considering the action arising from moving a
vortex around a closed loop. For orientation we first consider the related
purely bosonic problem of a vortex in a two boson condensate. The standard
bosonic Lagrangian density \cite{Popov} includes a 'non-dissipative time
derivative' term $i\gamma _{B}\psi ^{\dagger }\partial _{t}\psi $ with
coefficient $\gamma _{B}=1.$ A superfluid state is described by a condensate
amplitude $n_{s}=\left| <\psi >\right| ^{2}$ and condensate phase $\phi ,$
leading to a term $in_{S}\partial _{\tau }\phi $ in the action . This term
ensures that dragging a long straight vortex in the boson condensate around
a loop enclosing an area $s$ leads to a contribution to the action of $%
\Delta S=2\pi n_{S}s.$ This contribution is just the action appropriate to a
particle in a magnetic field of strength $h_{eff}=2\pi n_{S}$. For particles
on a lattice, a magnetic flux of $2\pi $ per unit cell has no dynamical
consequences, so that \ one measures $n_{S}$ modulo 1 per lattice site.
Finally, we consider the relation between the condesate density $n_{S}$ and
the total boson density $n_{B}$. Gauge invariance means that $-i\partial
_{t}\phi $ is a chemical potential, and therefore couples to the total
particle density, $n_{B}$. In simple boson problems, at $T=0$ the only
gapless excitation is the phase mode of the superfluid state and therefore $%
n_{S}=n_{B}$. As $T$ is increased from $T=0$ gapless 'normal fluid'
excitations occur. The presence or absence of Galilean invariance then
becomes crucial. One may interpret the $\partial _{t}\phi \,$\ term in the
vortex action in terms of the acceleration of a vortex in a given force. In
a Galilean invariant situation one expects that if a vortex is accelerated
it will drag all particles in the system with it, so that the coefficient of 
$\partial _{t}\phi $ is simply the total particle density in the system.
However, in a non-Galilean invariant system the superfluid component may
accelerate independently of the normal component and $n_{S}\leq n_{B}$.

We next consider a superconducting condensate made of paired electrons. The
important different here is that 'two-fluid' effects may be important even
at very low $T$ for example because of impurity-induced pairbreaking or of
vortices. The conventional result is that if a Landau expansion may be
constructed about a non-superconducting state (for example, very near to $%
T_{c}$ or in the presence of strong pair breaking) then one has 
\begin{eqnarray}
S_{sc} &=&\int dtd^{d}xN_{0}T_{0}^{2}  \label{ssc} \\
&&\left[ \gamma _{F}\frac{\Delta ^{+}\partial _{t}\Delta }{T_{0}^{3}}+\frac{%
\xi _{0}^{2}\left( \nabla \Delta \right) ^{2}}{T_{0}^{2}}+\frac{T-T_{c}}{%
T_{0}}\frac{\Delta ^{2}}{T_{0}^{2}}+u\frac{\Delta ^{4}}{T_{0}^{4}}\right]  
\nonumber
\end{eqnarray}
Here $N_{0}$ is the electronic density of states, $T_{0}$ is an energy scale
of of the order of the transition temperature or the pairbreaking scattering
rate $u$ is a coefficient of the order of unity and the coefficient $\gamma_F$
of the time derivative term has both real and imaginary parts,
$\gamma_F=\gamma_F'+i\gamma_F''$. In  a usual superconductor the dissipative
(real) part of $\gamma_F$ is of the order of unity (the conventional result is
$\gamma_F''=\pi/8$).  Because a time
dependent phase is a contribution to the chemical potential we may identify 
the imaginary part $\gamma _{F}''$ 
with $-\partial T_{c}/\partial \mu $. In conventional
superconductors this is very small (of order $T_{c}/E_{F}$) so the
total coefficient of the dissipationless time derivative,
$N_{0}T_{0}\gamma_{F}'' \sim \left( T_{c}/E_{F}\right) ^{2}$ 
is extremely small and for most
purposes may be neglected (for exceptions see e.g \cite
{Geshkenbein96,Dorsey91}). Of course in a conventional superconductor with
weak pairbreaking, a Landau expansion only applies for temperatures very near
to $T_{c}$ and as $T\rightarrow 0$ one expects 
$\gamma_F' \rightarrow 0$ while  the
dissipationless term $N_{0}T_{0}\gamma _{F}''$ must
approach the total fermion density $n$. For a type-II superconductor in a
magnetic field one similarly expects that $N_{0}T_{0}\gamma _{F}''\rightarrow n
$ only for \ temperatures of the order of the core state level spacing $%
T_{c}^{2}/E_{F}$ and only in the 'super-clean' limit. To summarize, known
results from simple fermion and boson problems imply that the
dissipationless time derivative term in the superfluid action involves a
non-universal coefficient which depends on the interplay between the
superfluid and non-superfluid components of the system, and is in general
quite small for fermion-based superfluids and is of the order of the
particle density for boson-based systems.

We now turn to the boson-fermion-gauge-field problem of interest here. The
discussion above shows that there is no simple, generally valid expression
for the coefficient, except in the $T\rightarrow 0$ no-pairbreaking limit,
in which the coefficient is the total particle density. Nevertheless a few
remarks can be made and limits can be estimated. At high dopings ($xt>J$)
the bosons condense (or quasi-condense) at temperatures well above $T_{c}$
so in this limit superconductivity arises out of a more or less fermi-liquid
like state, so near $T_{c}$ one expects the resulting superfluid state to be
described by a non-dissipative coefficient $\gamma $ which is of a fermionic
order of magnitude and thus much less than unity. We note, though, that in
the present model the spinon fermi energy is of order $J$ and the pairing
amplitude varies from a (not too small) fraction of $J$ at low doping to a
very small value at high doping \cite{Kotliar88} so $\partial \Delta
/\partial \mu \,\ $\ is not particularly small; the main smallness is
provided by the factor $N_{0}T_{0}\sim T_{c}/J$. \ On the other hand, as
doping is reduced the physics changes. The fermions pair at a higher scale
and the superconducting transition is set by the condensation scale of the
bosons. An upper bound on the non-dissipative coefficient is then set by the
total particle density (modulo $1$) i.e. $\gamma <x$. In this limit one
expects that by temperatures of the order of $T_{c}$ the fermions are mostly
paired, so that a Landau expansion is not appropriate and the coefficient is
set by bosonic physics; i.e. $\gamma $ is not that different from its $T=0$
no-pairbreaking value. We therefore propose the following approximate interpolation
formula for the efffecitve magnetic field $B_{eff}$ implied by the non-dissipative
terms in the action of our problem:

\begin{equation}
B_{eff}\approx \frac{1}{(\pi x)^{-1}+(\Delta /J)^{-1}}  \label{beff}
\end{equation}

We turn next to the dissipative term. In the gauge model, the dissipation
arises from the continuum of spinon excitations, and we assume that the
temperature, magnetic field or impurity density is large enough that an
appreciable number of these exist, and may be characterized by a spinon
conductivity $\sigma _{sp}$ which we take to be local on the scales of
interest. The 'electric field' felt by the spinons is\ $\partial _{t}\left( 
\frac{1}{2}\nabla \phi -a\right) $ so that the dissipative contribution to
the action is 
\begin{eqnarray}
S_{diss} &=&\frac{\sigma _{sp}}{2}\int d^{2}rdtdt^{\prime }  \label{Sdiss} \\
&&\partial _{t}\left( \frac{1}{2}\nabla \phi -a\right) _{t}K(t-t^{\prime
})\partial _{t}\left( \frac{1}{2}\nabla \phi -a\right) _{t^{\prime }} 
\nonumber
\end{eqnarray}
with $K$ the fourier transform of $1/\left| \omega \right| $. In a slowly
moving vortex with center position $\overrightarrow{X}_{v}(t)$, $\frac{1}{2}%
\nabla \phi -a$ is a function $\widetilde{a}(\overrightarrow{r}-%
\overrightarrow{X}_{v}(t))$ Substitution into Eq \ref{Sdiss} and some
rearrangement leads to the standard form for a particle moving in a
dissipative medium, namely 
\begin{equation}
S_{diss}=\frac{\eta }{2}\int dtdt^{\prime }\left( \partial _{t}%
\overrightarrow{X}_{v}(t)\right) K(t-t^{\prime })\left( \partial _{t^{\prime
}}\overrightarrow{X}_{v}(t^{\prime })\right)  \label{Sdiss2}
\end{equation}
with viscosity $\eta $ given by 
\begin{equation}
\eta =\frac{\sigma _{sp}}{2}\int d^{2}r\sum_{ij}\left( \frac{\partial \left( 
\frac{1}{2}\nabla \phi -a\right) _{i}}{\partial r_{j}}\right) ^{2}
\label{eta}
\end{equation}
Use of Eq \ref{jb} shows that $\rho _{F}\left( \frac{1}{2}\nabla \phi
-a\right) $ is just the physical current $\overrightarrow{j}_{phys}=\rho
_{B}f^{2}(r/\xi )\widehat{\theta }/r$. Substitution of our result for $f$
leads to (Note that the logarithm comes from the angular derivative ($\nabla
\rightarrow \frac{1}{r}\partial _{\theta })$) 
\begin{equation}
\eta =\frac{\pi \sigma _{sp}}{\xi ^{2}}f_{0}^{2}\left( \frac{\rho _{B}}{\rho
_{F}}\right) ^{2}\ln (\xi /\xi _{F})  \label{etafinal}
\end{equation}
We see that as the insulator is approached, the vortex viscosity vanishes
very rapidly, indeed as $x^{3}$. One factor of $x$ comes from the large size
of the vortex (proportional to $\xi ^{-2}$); the other two factors $\left( 
\frac{\rho _{B}}{\rho _{S}}\right) ^{2}$ come from the decreased coupling of
the vortex motion to the spinons; we interpret this as a signature of the
vanishing of the quasiparticle charge in this theory (for a discussion of
other signatures see \cite{Millis96,Ioffe01a}).

It is instructive to view this result in a slightly different way. The
'internal electric field' $e$ felt by the spinons is $\partial _{t}\left( a-%
\frac{1}{2}\nabla \phi _{F}\right) =\frac{1}{e\rho _{F}}\partial _{t}\left(
j_{phys}(r)\right) $. Application of the usual composition rules of the
gauge theory \cite{Ioffe89} shows that the physical electric field $E=e\rho
_{F}/\rho _{B}$ (in the limit $\rho _{F}>>\rho _{B}$) so that the physical
electric field $E$ generated by a moving vortex is 
\begin{eqnarray}
E(r)= &&\frac{1}{e\rho _{B}}\partial _{t}\left( j_{phys}(r)\right)  \nonumber
\\
= &&\partial _{t}\left( \overrightarrow{X}_{v}(t)\right) \cdot 
\overrightarrow{\nabla }\left( \frac{f^{2}(\frac{r-X_{V}}{\xi })\widehat{%
\theta }}{r-X_{V}}\right)  \label{Evort2}
\end{eqnarray}
This shows that a moving vortex generates an electric field which varies on
the length scale set by the physical current. \ $E(r)$ varies as $1/r^{2}$
far from the vortex, and as $1/r$ for $\xi _{F}<r<x^{-1/2}$. \ A standard
argument \cite{Tinkham90} says that the dissipative contribution may be
estimated by multiplying the square of the electric field by the physical
conductivity, which from Eq \ref{Hmixspinon2} is of order $\left( \rho
_{B}/\rho _{F}\right) ^{2}$, leading again to Eq \ref{etafinal} for $\eta $.

The result for $\eta $, along with the estimates of the core energy, has
implications for the width of the resistive superconducting transition in
this model. In a two dimensional superconductor, the resistive transition is
of the Kosterlitz-Thouless vortex unbinding type. Near to the transition
point, but in the normal state, one has a dilute gas of vortices and
antivortices. The physical conductivity is the sum of the conductivity due
to the moving vortices and the conductivity due to the quasiparticles. Use
of Eq \ref{Hmixspinon2} and Eq \ref{etafinal} leads to 
\begin{equation}
\sigma _{phys}=\left( \frac{\rho _{B}}{\rho _{F}}\right) ^{2}\sigma
_{sp}\left( 1+\frac{\pi f_{0}^{2}\ln (\xi _{B}/\xi _{F})}{n_{V}\xi ^{2}}%
\right)  \label{sigphys}
\end{equation}
This, in this model the conductivity is dominated by proximity to the
superconducting phase only if vortices are dilute on the length scale set by
the supercurrent pattern. However, even when vortices overlap from the
'current' point of view, they may still be dilute on the scale set by the
'quasiparticle core size' $\xi _{F}$, so we would therefore expect the gap
in the fermionic excitation spectrum to persist to much higher temperatures
than does the 'superfluid' contribution to the conductivity.

The density (and nature) of thermally excited vortices bears further
discussion. The factors influencing the density of vortices may be
understood from simple mean field free energy arguments. For conventional $%
(h/2e)$ vortices the mean-field free energy as a function of conventional
vortex density $n_1$ is (note that the position of the conventional vortex
may be defined to within $\xi_F$ but the logarithm from the superfluid
stiffness is cut off by $\xi$)

\begin{equation}
E_{n=1,m=0}=E_{core}n_{1}-\rho _{S}n_{1}ln(n_{1}\xi ^{2})-Tn_{1}ln(n_{1}\xi
_{F}^{2})
\end{equation}

Away from the K-T fluctuation regime, $\rho_S$ becomes less than $T$ and we
have, approximately,

\begin{equation}
n_1\xi^2=\frac{\xi^2}{\xi_F ^2}e^{-E_{core}/T}  \label{n1}
\end{equation}

The factor $=\frac{\xi^2}{\xi_F ^2}$ makes it easier than one might guess to
obtain a reasonable density of conventional vortices; of course the small
value of the core energy, arising as discussed above from the proximity in
energy of other gapped states within the d-RVB theory, is also important.

It is also of interest to consider the density $n_{2}$ of doubly quantized
vortices, (i.e. vortices in the boson field only), because their motion
leads to dissipation too. As noted above, in this case the core energy is
entirely determined by the superfluid stiffness so we have ($\Phi $ is a
scaling function which is exponentially small at large argument and becomes
of order unity when its argument becomes of order unity and the factor of 4
comes from the doubly quantized nature of the vortex)

\begin{equation}
n_2 \xi^2=\Phi(4\rho_S/T)  \label{n2}
\end{equation}

Therefore, even if the core energy of a conventional vortex is very high,
(which, in the d-RVB theory it is not) above a temperature scale set by $%
\rho_S$ double quantized vortices will proliferate and will suppress the
superfluid contribution to the conductivity.

\section{Quantal fluctuations and the melting of the vortex lattice.}

We now consider the physics at low temperature in an applied magnetic field.
An applied field induces a vortex lattice and, ultimately, a
non-superconducting state. In a conventional (BCS) superconductor the
transition is driven by the collapse of the superconducting gap. In the
present model the important physics involves quantal fluctuations in the
positions of the vortices, leading via a first order transition to a
'quantal vortex liquid' state. \ In particular, although an applied field
leads to pairbreaking, and thus to a non-vanishing density of quasiparticles
and to a reduction of the superfluid stiffness and gap amplitude, in the
gauge model of interest here these effects are small (order $x^{-2}$) \cite
{Ioffe01a} so we neglect them.

We estimate the magnitude of the quantal fluctuations of the vortices by
considering the fluctuation in position of one vortex about its ideal
Abrikosov lattice position.To do this we note that in in a vortex lattice
the force term in Eq \ref{xv} may be written for small amplitude
displacements as $F_{V}=K_{latt}X_{V}$ \ where $K_{latt}$ is the coefficient
of the quadratic term in the restoring potential acting on the vortex and
arising from the other vortices in the vortex lattice (plus any pinning
forces which may exist, and which we do not treat) . Eq \ref{xv} applies
only for frequencies less than a cutoff frequency which is the minimum of $%
\Delta $ and the boson frequency scale $xt$. $K_{latt}$ is of the order of
the physical superfluid stiffness $\rho _{B}$ divided by the square of the
intervortex spacing, i.e. $K_{latt}=K_{0}n_{V}\rho _{B}$. The long range
(logarithmic) form of the intervortex potential means that $K_{0}<<1$.We may
now quantize Eq \ref{xv} and thereby estimate the zero point fluctuations of
a vortex as ($\omega _{n}$ is a Matusbara frequency and $\sigma^y$ is a
Pauli matrix) 
\begin{equation}
<X_{V}^{2}>=Tr\sum_{n}\left[ \eta \left| \omega _{n}\right| +\sigma
^{y}B_{eff}\omega _{n}+K\right] ^{-1}  \label{rms}
\end{equation}
We estimate that the vortex lattice melts when the mean square vortex
displacement divided by the square of the intervortex distance $b$, i.e. $%
<X_{V}^{2}>/b^{2},$ becomes of the order of the square of the Lindemann
number $c_{L}^{2}$. The value of the Lindemann number depends on the
physical situation and varies from $c_{L}^{2}=0.01$ to $c_{L}^{2}=0.1$ \cite
{Lindemann}. The estimate of $<X_{v}^{2}>$ depends on the values of $\eta $
and $B_{eff}$ and in particular on whether the system is in the super-dilute
regime ($B_{eff}\sim \pi n_{B}$ and $\eta \rightarrow 0$) or the dilute
regime ($B_{eff}$ unimportant; $\eta $ dominant). In the super-dilute regime
we neglect the $\eta $ and find (the numerical factors give the relation
between $b^{2}$ and $n_{V}$ for a triangular vortex lattice) 
\begin{equation}
\frac{<X_{V}^{2}>}{b^{2}}=\frac{3\sqrt{3}n_{V}}{8B_{eff}}
\label{superdilute}
\end{equation}

In fact, if dissipation is negligible then we may look at the problem in a
different way. The vortices move in an effective field which is large,
leading to Landau level quantization. If the density of vortices is much
less than the density of bosons, $n_{V}<<n_{B}$ then only the lowest Landau
level is populated and we may use known results for the melting of the
Wigner crystal in a high magnetic field to argue that lattice melting occurs
when $n_{V}/n_{B}\approx 1/10$ corresponding to a Lindemann number $%
c_{L}^{2}\approx 0.01$ reasonably consistent with known results for two
dimensional triangular lattices \cite{Lindemann}.

In the dilute regime, the frequency cutoff is needed. If $\rho _{B}>\Delta $
then this scale is $\Delta $ and we obtain 
\begin{equation}
\frac{<X_{V}^{2}>}{b^{2}}=\frac{3\sqrt{3}n_{V}}{4\pi \eta }\ln \left[ 1+%
\frac{\Delta \eta }{K}\right]  \label{dilute}
\end{equation}

Use of Eq \ref{etafinal} and our estimate for $K$ shows that the important
dimensionless parameter is given by 
\begin{equation}
\frac{\Delta \eta }{K}\sim \frac{\Delta }{\rho _{B}}\left( \frac{\xi _{V}^{2}%
}{\xi _{B}^{2}}\right) \frac{\pi \sigma _{sp}}{K_{0}}\left( \frac{\rho _{B}}{%
\rho _{F}}\right) ^{2}\ln \left( \frac{\xi _{B}}{\xi _{F}}\right)
\end{equation}
This parameter may be larger or smaller than unity, because $\rho _{B}/\rho
_{S}\sim x$ while $K_{0}<<1$ and $\sigma _{sp}$ must by consistency be of
order $1/x$ (because we have written clean-limit formulae which require that 
$\sigma _{sp}=p_{F}l_{F}/2\pi >p_{F}\xi \,_{F}/2\pi >\rho _{S}/\rho _{B}\sim
1/x$. However, as one goes more deeply into the small $\Delta $ regime the
parameter shrinks and we obtain 
\begin{equation}
\frac{<X_{V}^{2}>}{b^{2}}=\frac{3\sqrt{3}\Delta }{4\pi K_{0}\rho _{B}}\left( 
\frac{\xi _{V}^{2}}{\xi _{B}^{2}}\right)
\end{equation}
Thus in the dilute limit the lattice still melts when the flux per boson is
of the order of the square of the Lindemann number , up to a factor of order 
$\rho _{B}/\Delta $.

The melted phase is an interesting example of a 'non-fermi-liquid': the
vortex motion is damped (albeit weakly) so the model is characterized by a
non-vanishing $\sigma _{xx}$ and $\sigma _{xy}$, as in a normal metal,
(although the $\sigma _{xy}$ value is rather large) but also by a
'non-Luttinger' fermi surface with a gap over large regions of the nominal
fermi surface.

\section{Conclusion}

One of the most interesting aspects of high temperature superconductivity is
the 'pseudogap' regime of underdoped materials. This regime is characterized
by a gap (of approximately d-wave form) in the quasiparticle spectrum but
neither long ranged superconducting order nor particularly noticeable
superconducting fluctuations. One possibility is that this regime involves
electron pairing (as in a conventional superconductor) but with long ranged
superconducting order disrupted by strong phase fluctuations, arising
physically from the strong suppression of charge response expected mear a
Mott insulator. A difficulty with this idea is the absence of noticeable
'paraconductivity': transport measurements indicate a critical regime of
order $10K$ at most \cite{Ong93,Corson98}.

In this paper we studied one theoretical implementation of the 'phase
fluctuation' scenario for the pseudogap, namely the d-RVB regime of the $%
U(1) $ gauge theory \ of lightly doped Mott insulators. In this approach the
spin degrees of freedom are mostly paired into a d-wave pairing state and
the low $T$ charge response is essentially that of a superfluid, but with
properties strongly affected by proximity to the Mott transition. One
important feature of the model is that the vortex excitations are
characterized by two length scales: the 'quasiparticle coherence length' $%
\xi _{F}=v/\Delta $ which controls the distance over which the excitation
spectrum differs from that far from a vortex, and the 'current coherence
length' $\xi $ which varies as the square root of the doping and cuts off
the familiar $1/r$ divergence of the supercurrent near a vortex. We studied
the charge transport properties (many of which are dominated by vortices)
and showed in particular that the electric field created by a moving vortex,
the dissipation due to moving vortex, the value of $H_{c2}$ and the size of
the fluctuation regime near the resistive transition are all controlled by $%
\xi $ which diverges near the Mott transition, rather than by the
'quasiparticle length $\xi_F$ which does not.

As noted elsewhere \cite{Lee97,Millis96,Ioffe01a} the theory disagrees in a
number of ways with experiments; the most significant difficulty is the
small value and strong doping dependence of the 'quasiparticle charge'
defined in Eq \ref{Hmixspinon2}. Further, our calculation has a number of
phenomenological aspects. For example, we assumed a finite 'spinon
conductivity' which could reasonably be expected to arise from the 'gapless
fermi arcs' induced by a non-vanishing temperature or applied magnetic
field, but we did not attempt to calculated this from first principles, nor
did we investigate the subtle quantum mechanics of fermions in the presence
of conventional vortices. However, we believe the results presented here are
useful because they provide an explicit demonstration in a well-defined
model that if the supercurrent-defined correlation length is parametrically
larger than the quasiparticle-defined length, then the resistive properties
are controlled by the length scale over which the supercurrent varies.

Other workers have observed that the theory admits doubly quantized
vortices. We have noted that they proliferate above a scale defined by the
physical superfluid stiffness and (if the core energy of conventional
vortices were large, which it is not in this model) would suppress the
superconducting fluctuation contribution to the conductivity. We also showed
how, in this model, the 'non-dissipative time derivative', whose importance
was stressed in \cite{Geshkenbein96}, is important for the estimation of the
upper critical field, and argued that it crosses over from the 'fermionic'
value $T_{c}/E_{F}$ to the 'bosonic' value proportional to the density of
charges $x$. A subsequent paper will apply these ideas to a different model
of high-T$_{c}$ superconductivity.

{\it Acknowlegdgement:} This work was supported by NSF-DMR-00081075 and the
Institute for Theoretical Physics at Santa Barbara. AJM thanks N. P. Ong, Y.
B. Kim, D. T. Son, M. P. A. Fisher and especially P.A. Lee for helpful
conversations.

\section{Appendix: Fermi-liquid-based approaches to doped Mott insulators}

This Appendix treats the case of superconductivity developing out of a state
which is well described by the usual fermi liquid theory. The necessary
formalism was developed by Larkin and Leggett , and some of the results were
sketched elsewhere \cite{Landaurefs}. A fermi-liquid state is characterized
by a quasiparticle velocity $v^{\ast }(\theta )$ which may depend on
position ($\theta $) on the fermi surface, a characteristic energy scale $%
E^{\ast }$ and a Landau interaction function $T(\theta ,\theta ^{\prime })$.
In order for fermi liquid theory to be applicable, the maximum
superconducting gap $\Delta _{0}$ must be less than $E^{\ast }$.
Transcription of the standard results to the langauge of the section above
leads to 
\begin{eqnarray}
\rho _{s0} &=&<v^{\ast }(\theta )\left( 1-T\right) _{\theta ,\theta ^{\prime
}}^{-1}v^{\ast }(\theta ^{\prime })>  \label{rhosfl} \\
v_{F} &=&v^{\ast }  \label{vffl} \\
\frac{d\rho _{S}}{dT} &=&\frac{<v^{\ast }(\theta )\left( 1-T\right) _{\theta
,\theta _{1}}^{-1}L(\theta _{1})\left( 1-T\right) _{\theta _{1},\theta
^{\prime }}^{-1}v^{\ast }(\theta ^{\prime })>}{<v^{\ast }(\theta )^{2}>}
\label{Z2fl}
\end{eqnarray}
where the angle backet means multiplication by density of states and average
over the fermi line and $L(\theta )=\int \frac{d\omega d\theta }{2\pi ^{2}}%
\frac{\omega }{2T^{2}\coth (\frac{\omega }{2T})}\frac{\Delta (\theta )^{2}}{%
\left( \omega ^{2}+\Delta (\theta )^{2}\right) ^{3/2}}.$

Comparison of the these results with those presented in section $II$ shows
that the $U(1)$ theory corresponds to a fermi liquid with a weakly
angle-dependent Landau interaction function whose 'current-channel' value is
of order $1/x$ \cite{Landaurefs}.

We now consider the situation in more detail by calculating the low-T
current-current correlation function for a superconducting fermi liquid,
making the usual assumption that the Landau interaction function may be
decomposed into angular channels in the conventional way, and that the
maximum value of the superconducting gap is small compared to the
characteristic quasiparticle energy scale $E^{\ast }$ so that quasiparticle
damping effects may be neglected. The gauge-invariant current-current
correlation function is then \cite{Landaurefs} 
\begin{equation}
\chi _{jj}(q,\Delta )=\frac{\chi _{qp}(q,\Delta )}{1+I_{1}\chi
_{qp}(q,\Delta )}  \label{chifl}
\end{equation}
with $I_{1}$ the current-channel Landau interaction parameter (so that the
Landau parameter $F_{1s}=2I_{1}\chi _{qp}(0,\Delta )$) and 
\begin{eqnarray}
\chi _{qp}(q,\Delta ) &=&T\sum_{\omega }\int \frac{d\theta }{\left( 2\pi
\right) ^{2}v^{\ast }(\theta )}v_{x}^{\ast 2}\frac{\Delta (\theta )^{2}}{%
\sqrt{\Delta (\theta )^{2}+\omega ^{2}}}  \label{chiqp} \\
&&\frac{1}{\left( \omega ^{2}+\Delta (\theta )^{2}+\left( v^{\ast }q\cos
(\theta -\theta _{q})\right) ^{2}/4\right) }  \nonumber
\end{eqnarray}
(Note that the result depends on the angle $\theta _{q}$ between the
direction of $q$ and the nodes in the gap).

Standard calculations \cite{AGD} show that at $vq>>\Delta _{0}$, $\chi
_{qp}\sim \Delta _{0}/vq$ so that we expect an appreciable change in $\chi
_{jj}$ when $vq\sim \Delta _{0}/x$. In other words, a naive application of
fermi liquid theory would predict a very short characteristic length scale
of order $xv/\Delta $ . However, from the usual physical picture of the
doped Mott insulator as a dilute collection of holes one might expect the
interparticle spacing $x^{-1/2}$ to be an important scale. \ The discrepancy
is resolved by noting that the Landau parameter presumably varies on the
scale $x^{-1/2}$. This effect is beyond the scope of Landau theory but is
captured in the $U(1)$ approach.

\end{document}